\UseRawInputEncoding
\documentclass[aps,prc,twocolumn,showkeys,showpacs]{revtex4}
\usepackage{graphicx}
\usepackage{dcolumn}
\usepackage{color}
\usepackage{bm}

\begin{document}

\newcommand{\req}[1]{(\ref{#1})}
\newcommand{\bel}[1]{\begin{equation}\label{#1}}
\newcommand{\belar}[1]{\begin{eqnarray}\label{#1}}
\def\eps{\epsilon}
\def\mev{\;{\rm MeV}}
\def\tc{\textcolor{red}}
\def\tb{\textcolor{blue}}
\def\pr{\prime}
\def\dpr{\prime\prime}

\title{Memory effects in Langevin approach to nuclear fission process}

\author{ F. A. Ivanyuk}
\email{ivanyuk@kinr.kiev.ua}
\affiliation{Institute for Nuclear Research, 03028 Kiev, Ukraine}

\author{ S. V. Radionov}
\email{sergey.radionov18@gmail.com}
\affiliation{Institute for Nuclear Research, 03028 Kiev, Ukraine}

\author{C. Ishizuka}
\email{chikako@nr.titech.ac.jp}
\affiliation{Tokyo Institute of Technology, Tokyo, 152-8550 Japan}
\author{S. Chiba}
\email{chiba.satoshi@nr.titech.ac.jp}
\affiliation{Tokyo Institute of Technology, Tokyo, 152-8550 Japan}

\date{today}

\begin{abstract}
We present the  calculations within the schematic Langevin approach to investigate
the dependence of fission width on the memory time and the excitation energy at low temperatures where the quantum fluctuations play an important role. For this, we consider the simple one-dimensional model with the potential energy given by the two parabolic potentials (Kramers potential). For the friction and mass parameters, we use the deformation independent values fitted to the results obtained earlier within the microscopic linear response theory. We have found out that at small excitation energies (comparable with the fission barrier height) the memory effects in the friction and random force acts on the fission width in opposite direction. The total effect depends very much on the value of damping parameter $\eta$. In the low viscosity region the fission width $\Gamma_f$ {\it grows} as function of $\eta$ and {\it decreases} as function of $\tau$. In the high viscosity region, the tendency is opposite.  The fission width $\Gamma_f$ {\it decreases} as function of $\eta$ and {\it increases} as function of $\tau$.  Such dependence is common both for small and large excitation energies.
\end{abstract}

\maketitle

\section{Introduction}
The Langevin approach is applied to the description of the nuclear fission process already for few decades \cite{nix76,wada93,froebrich,pomorski1996,asano,adeev2005,aritomo,mazurek,our17,sierk,scirep,kosenko}. In these works the one-five dimensional Langevin equations were solved with macroscopic or microscopic \cite{hofrep} transport coefficients. The approach describes quite successfully the mass distributions and kinetic energies of fission fragments, the multiplicities of emitted neutrons. Still, there are few long-standing problems that are not described properly up to now. One of such phenomenons is
the rapid transition of fission fragments mass distribution from symmetric to mass asymmetric in spontaneous fission of Fermium isotopes \cite{lane96}. As it was shown by Flynn {\it et al}  \cite{Flynn_1975}, the mass distributions of fission fragments of Fermium isotopes are very sensitive to the excitation energy. The mass distributions in spontaneous fission and thermal neutron-induced fission look very different.

This circumstance indicates the necessity to have an accurate description of the fission process at extremely low excitation energies, below the fission barrier. At low excitation energies, the quantum effects in transport coefficients and the memory effects in friction and random force become important.

Below in this work, we try to investigate the influence of memory effects on the fission width at low excitation energies. To make it extremely clear we consider the simple one-dimensional model with the potential energy given by two parabolic potentials (Kramers potential). For the friction and mass parameters, we use the deformation independent values fitted to these obtained within the microscopic linear response theory \cite{hiry}.


In Section II we check how our one-dimensional numerical code reproduces the known analytical expression for the decay width. In Section III the non-Markovian Langevin equations are solved at constant temperature and the dependence of the fission width on the value of relaxation time, the damping parameter, and the excitation energy is clarified. In Section IV the role of effective temperature is investigated.
Section V contains the summary.

\section{Classical Langevin equations}
In order to check the numerical code we will solve first the one-dimensional Langevin equations with the "white" noise,
\belar{lange1}
dq/dt&=&p(t)/M ,\nonumber\\
dp/dt&=&-\partial E_{pot}/\partial q - \gamma p /M + \sqrt{D}\,\xi,
\end{eqnarray}
where $\xi$ are the normally distributed random numbers with the properties
\bel{rand}
\langle\xi(t)\rangle=0,\quad \langle\xi(t)\xi(s)\rangle=2 \delta(t-s)
\end{equation}
and the diffusion coefficient $D$ is given by Einstein relation, $D=\gamma T$. The temperature here is considered to be time-independent parameter, related to the excitation energy $E^*$ by the Fermi-gas relation $E^* = a T^2$, and for the level density parameter $a$ we use the approximation \cite{wada93},
\bel{denspar}
      a =A/14.61(1.0+3.114 A^{1/3}+5.626 A^{2/3}), 
\end{equation}
where $A$ is the mass number.

For the potential energy,  we choose the simplest two-parabolic (Kramers) potential, see Fig.~\ref{poten},
\belar{epot}
{\rm E}_{pot}(q)&=&(2V_b/q_0^2)q(q-q_0),~~0<q<q_0,
\\
&=&(2V_b/q_0^2)(q-q_0)(2q_0-q),~~q_0<q<2q_0.\nonumber
\label{epota}
\end{eqnarray}
This potential depends on two parameters, the barrier height $V_b$,  and the barrier width $q_0$. We have fixed the barrier height as $V_b=6 \mev$, which is close to the value of the fission barrier of actinide nuclei. The width of the barrier is somewhat uncertain. It depends on the definition of the collective coordinate $q$ and the model for the potential energy. For simplicity, we have put here $q_0=1.0$. In principle, one should check the dependence of final results on $q_0$.

For the potential \req{epot} one can define the stiffness $C=d^2 E_{pot}/dq^2=4V_b/q_0^2$ and the frequency of harmonic vibrations $\omega_0=\sqrt{C/M}$. In present work we fix $\hbar\omega_0=1 \mev$, what is close to the frequency of collective vibrations calculated for $^{224}$Th in \cite{hiry} within the microscopic linear response theory.
\begin{figure}[htp]
\includegraphics[width=0.35\textwidth]{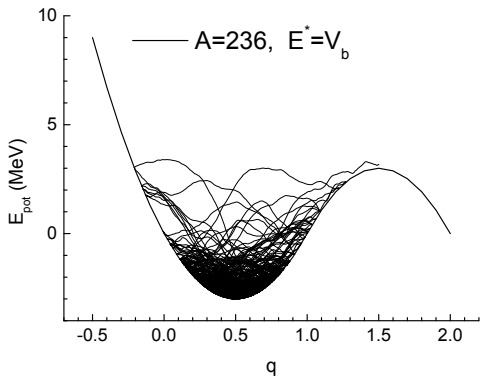}
\centering
\caption{The two-parabolic potential \protect\req{epot} and few examples of dynamical trajectories.}
\label{poten}
\end{figure}

Thus, we will have for the mass parameter the deformation and temperature-independent value
\bel{massa}
M=4 V_b / (\omega_0^2q_0^2) ,
\end{equation}
For the remaining parameter, friction coefficient $\gamma$, we will use a slightly modified approximation of \cite{hiry}
\bel{fric}
\gamma / M =0.6 (T^2+\hbar^2\omega_0^2/\pi^2)/(1+T^2/40)
\end{equation}
(here T is in MeV).
Now all the parameters of Eqs. \req{lange1} are fixed.

For the integration of Eqs. \req{lange1} we will need the integral $\int_t^{t+\Delta t} \xi(s) ds$, which is a sum of Gaussian random numbers,
and thereby is itself a Gaussian random number. Its average and variance can be calculated with
the statistical properties of $\xi(t)$, as
\belar{omega1}
\int_t^{t+\Delta t} \langle \xi(s)\rangle ds&=&0\,,\nonumber\\
\int_t^{t+\Delta t}\int_t^{t+\Delta t} \langle \xi(s)\xi(s^{\pr})\rangle dsds^{\pr}&=&2\Delta t\,.
\end{eqnarray}
Thus, we can describe $\int_t^{t+\Delta t} \xi(s) ds$, by a new Gaussian random number $\omega_1$, \bel{omega2}
\int_t^{t+\Delta t} \xi(s) ds=\omega_1\sqrt{\Delta t},
\end{equation}
such that $\omega_1$ have the properties
\bel{omega3}
\langle \omega_1\rangle=0,\qquad \langle \omega_1^2\rangle= 2.
\end{equation}
The random numbers $\omega_1$ can be constructed by Box-Muller transform \cite{box58}.

For the initial values of $q$ and $p$ we chose $q_{in}=q_0/2,  p_{in}=0$.
I.e., we start from the ground state deformation and assume that all the excitation energy is put into the intrinsic degrees of freedom.
For the integration of \req{lange1} we use the approximation schema,
\belar{pt5}
p(t+\Delta t)&=&p(t)-\bigg[\frac{\partial E_{pot}}{\partial q} + \gamma p(t) /M\bigg] \Delta t + \sqrt{ D \Delta t } \,\omega_1\nonumber\\
q(t+\Delta t)&=&q(t)+[p(t+\Delta t)+p(t)]\Delta t/2M .
\end{eqnarray}

In what follows we will be interested in the fission width of the system bound by the potential \req{poten}.
The fission width $\Gamma_f$ is defined assuming the exponential decay in time of the number of  "particles" in the potential well
\bel{decay}
P(t)=e^{-\Gamma_f t/\hbar}
\end{equation}
From here we will get
\bel{Gammaf}
\Gamma_f=-\hbar \ln[P(t)]/t
\end{equation}
By solving the Langevin equations one will get the set of time moments
$t_b$, at which some trajectories would cross the barrier, namely reach the value $q=2 q_0$. The total probability to get out of the potential well will be equal to
\bel{Pb}
P_b(t)=\sum_{t_b}\Theta(t-t_b)/N_{tr},
\end{equation}
with
\belar{theta}
\Theta(x)=\left\{
\begin{array}{rc}
&0,\,\text{if}\,x<0\,,\\
&1,\text{if}\, x \ge 0 \,.
\end{array} \right.
\end{eqnarray}
where $N_{tr}$ is the number of all trajectories taken into account.
Since the sum of probabilities to stay or get out of the potential well should be equal to unity  $P(t)+P_b(t)=1$, for $P(t)$ one gets
\bel{pt}
P(t)=1-\sum_{t_b}\Theta(t-t_b)/N_{tr}
\end{equation}
The fission width $\Gamma_f$ is defined then by the fit of $-\ln P(t)$ by linear in $t$ function
\bel{minim}
\int [-\ln P(t)-\Gamma_f t /\hbar]^2 dt - \text{min},
\end{equation}
From \req{minim} one gets
\bel{Gf}
\Gamma_f=-\hbar\int_0^{t_{max}} t \ln P(t) dt \,/ \int_0^{t_{max}} t^2 dt
\end{equation}
A demonstration of Eq. \req{Gf} is shown in Fig.~\ref{fig-pt}.
\begin{figure}[htp]
\includegraphics[width=0.35\textwidth]{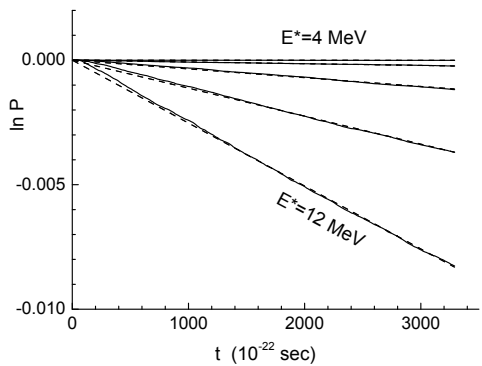}
\centering
\caption{The probability \req{pt} of "particle" to stay within the potential well (solid) and its fit by the linear function (dash).}
\label{fig-pt}
\end{figure}

The fission width $\Gamma_f$ calculated by Eqs. (\ref{pt5}, \ref{Gf}) is shown in Fig.~\ref{Fig_GammaK} as function of the excitation energy $E^*$.
For the comparison we show also the Kramers decay width $\Gamma_{K}$,
\bel{GammaK}
\Gamma_{K}=\frac{\hbar\omega_0}{2\pi}e^{-V_b/T}(\sqrt{1+\eta^2}-\eta)\,, \text{with}\,\,\, \eta\equiv\gamma/2M\omega_0\,,
\end{equation}
and the quantity inversely proportional to the so called "mean first passage time" \cite{vankampen, gardiner, risken}, $\Gamma_{mfpt}=\hbar/\tau_{mfpt}$ (dashed line in Fig.~\ref{Fig_GammaK}), where
\bel{tmfpt}
\tau_{mfpt}=\frac{\gamma}{T}\int_{q_1}^{q_2}d q\, e^{E_{pot}(q)/T}\int_{-\infty}^{q}d q^{\pr} e^{-E_{pot}(q^{\pr})/T},
\end{equation}
and $q_1$ and $q_2$ are the solutions of the equation $E_{pot}(q)=E^*$.
\begin{figure}[htp]
\centering
\includegraphics[width=0.4\textwidth]{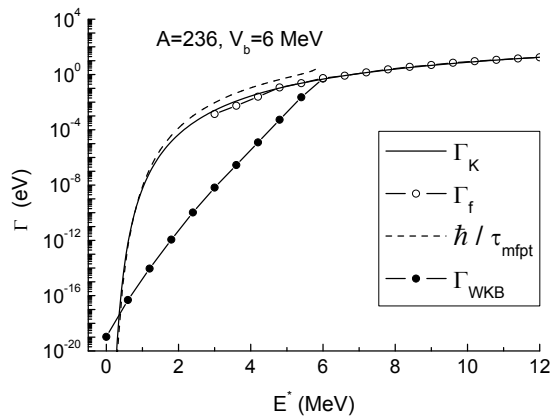}
\caption{The fission width \req{Gammaf}, as the solution of Eqs.~(\ref{lange1}-\ref{pt5}) (open dots), Kramers approximation \req{GammaK} (solid line), and WKB-approximation, Eqs.~(\ref{rwkb},\ref{pcoll}) (filled dots). Dash line - the inverse mean first passage time, see \req{tmfpt} below.
Parameters $N_{tr}=10000, t_{max}=3\cdot 10^{-18}$ sec.}
\label{Fig_GammaK}
\end{figure}
As one can see, above $E^*=3 \mev$ the fission width \req{Gammaf} is very close to Kramers approximation. The inverse mean first passage time is slightly larger. For smaller excitation energies $\Gamma_f$ is very small and the computations are too time-consuming.

The solid line marked by filled dots in Fig.~\ref{Fig_GammaK} is the decay width calculated within WKB-approximation, $\Gamma_{WKB}=\hbar ~r_{WKB}$, where the decay rate $r_{WKB}$ is given by \cite{landau5},
\bel{rwkb}
r_{WKB}=\frac{\omega_0}{2\pi}\exp{\left[-\frac{2}{\hbar}\int_{q_1}^{q_2}p(q) dq\right]}
\end{equation}
For the collective momentum $p(q)$ we used
\bel{pcoll}
p^2(q)=2(E^*-E_{pot})/M.
\end{equation}
As one can see, the quantum tunneling is dominant at rather small excitation energies, $E^*\leq 0.6 \mev$.

To check how our Langevin code works at large excitation energies, we have compared our numerical results with the calculations in \cite{scheuter83} within the Fokker-Planck approach. The $\Gamma_f$ shown in Fig.~\ref{Scheuter} is almost identical to Fig. 6 of \cite{scheuter83}.

It is important  to note that for small damping, below $\eta\approx 0.1$, the fission width $\Gamma_f$ is much smaller than the Kramers high viscosity limit \req{GammaK}, and approaches the low viscosity limit of Kramers \req{gammalv}
\bel{gammalv}
\Gamma_{LV}=\frac{\hbar\gamma}{M}\frac{V_b}{T}e^{-V_b/T}\,.
\end{equation}
\begin{figure}[htp]
\centering
\includegraphics[width=0.4\textwidth]{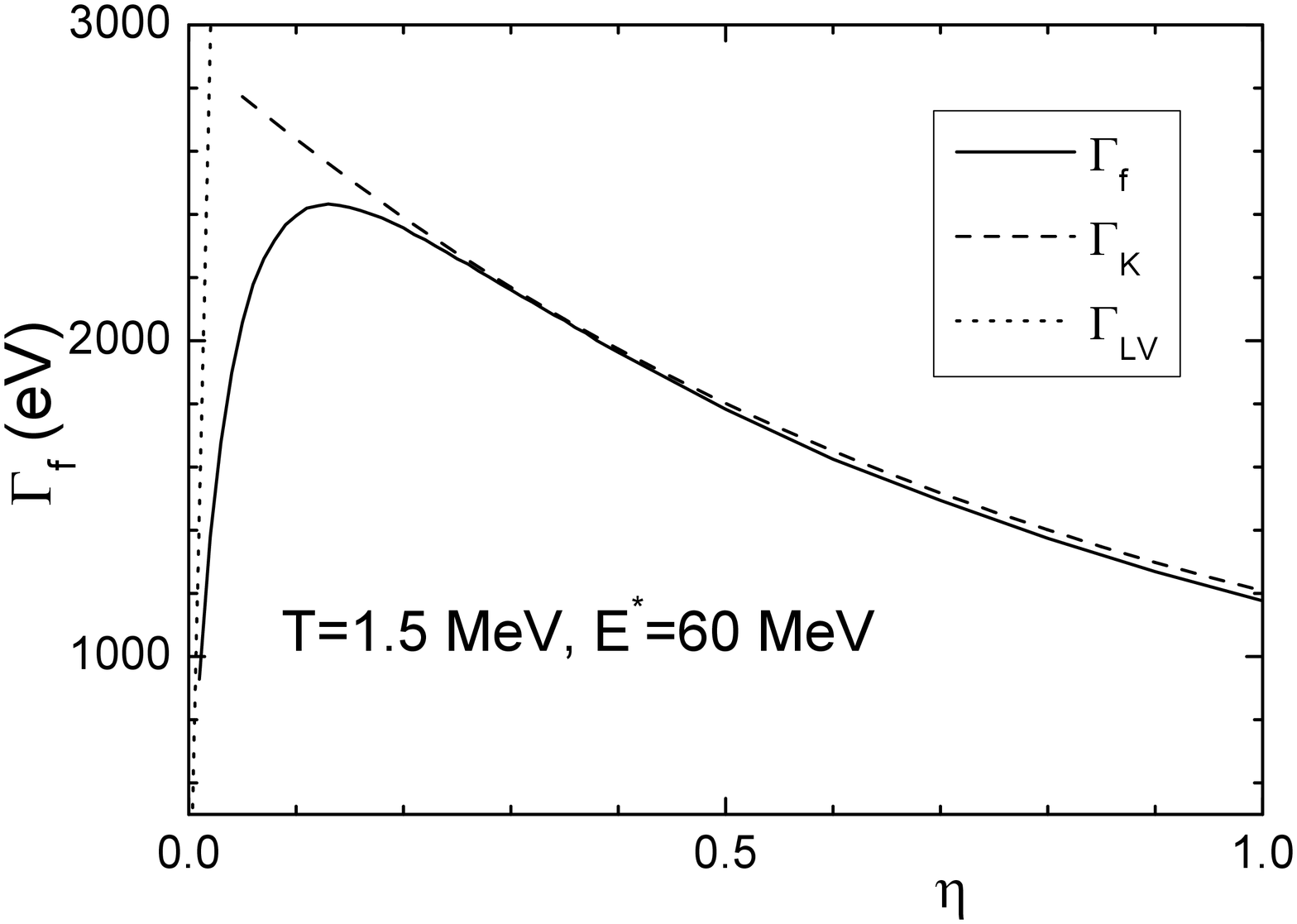}
\caption{The fission width \req{Gammaf}, as the solution of Eqs.~(\ref{lange1}-\ref{pt5}) (solid), the Kramers approximation \req{GammaK} (dash), and the low viscosity Kramers limit \req{gammalv} (dots). Parameters of calculations:
$\Delta t=5\cdot 10^{-24} sec, t_{max}=3\cdot  10^{-20} sec, N_{tr}=500 000$.
}
\label{Scheuter}
\end{figure}
\section{Non-Markovian Langevin equations}
The estimates of the memory effects on the nuclear dynamics are rather contradictive.
 In previous publications, one can find
both the statement that non-Markovian effects have
a substantial influence on the fusion or fission
processes \cite{korash,aick2004,washiyama} and the statement that non-Markovian
effects are very small \cite{froebrich}.

To clarify this question, we consider the non-Mar\-ko\-vi\-an Langevin equations that contain the memory effects, see \cite{abe-san}, Eqs. 376-377,
\belar{lange2}
dq/dt&=&p(t)/M ,\\
dp/dt&=&-\partial E_{pot}/\partial q - \int_{-\infty}^t\beta(t-t^{\pr})p(t^{\pr})dt^{\pr}+ \zeta(t),\nonumber
\end{eqnarray}
with $\beta(t-t^{\pr})=(\gamma/M)\exp{(-|t-t^{\pr}|/\tau)}/\tau$.

In the present work we will use a particular type of random numbers $\zeta(t)$ , that satisfy the equation
\bel{zetat}
d\zeta(t)/dt=- \zeta(t) / \tau +\xi/\tau\,
\end{equation}
and are used by the description of the so called Ornstein-Uhlenbeck processes.

The formal solution of Eq.\req{zetat} is
\bel{zetat1}
\zeta(t)=\zeta(t_0)e^{-\frac{t-t_0}{\tau}}+\frac{1}{\tau}\int_{t_0}^t e^{\frac{s-t}{\tau}}\xi(s)ds.
\end{equation}
Putting $t_0=-\infty$ one gets
\bel{zetat2}
\zeta(t)=\frac{1}{\tau}\int_{-\infty}^t e^{\frac{s-t}{\tau}}\xi(s)ds.
\end{equation}
With help of Eq. \req{zetat2} one can easily gets the correlation $\langle\zeta(t)\zeta(t^{\pr})\rangle$,
\belar{zeta3}
\langle\zeta(t)\zeta(t^{\pr})\rangle&=&\frac{1}{\tau^2}\int_{-\infty}^tds e^{\frac{s-t}{\tau}}\int_{-\infty}^{t^{\pr}} ds^{\pr} e^{\frac{s^{\pr}-t^{\pr}}{\tau}}\langle\xi(s)\xi(s^{\pr})\rangle \nonumber\\ 
&=&e^{-{\frac{|t-t^{\pr}|}\tau}}\slash \tau.
\end{eqnarray}
Here parameter $\tau$ characterize the strength of the memory effects.
The correlation function \req{zeta3} was used earlier in \cite{abe-san,adeev2008,kora2009}. The more complicated form of the correlation function was derived in \cite{aick2005,kora2010}. A new approach to treating the coupling between the Hamiltonian of the system and the environment (a bath of harmonic oscillators) was presented recently in \cite{tokieda}.

Here we will use the simplest form \req{zeta3}. In this case for the Langevin equations we will have
\belar{lange3}
dq/dt=p(t)/M ,\qquad\qquad\qquad\qquad\qquad\qquad\\
\frac{dp}{dt}=-\frac{\partial E_{pot}}{\partial q} - \frac{1}{\tau}\int_0^t dt^{\pr}e^{-\frac{t-t^{\pr}}{\tau}}\gamma p(t^{\pr}) /M + \sqrt{D}\,\zeta,\nonumber
\end{eqnarray}
Now we will introduce the notation $R(t)$ for the retarded friction,
\bel{roft}
R(t)\equiv\frac{1}{\tau}\int_0^t dt^{\pr}e^{-\frac{t-t^{\pr}}{\tau}}\gamma p(t^{\pr}) /M\,.
\end{equation}
It can be easily checked that $R(t)$ obeys the differential equation
\bel{drdt}
\frac{dR}{dt}=-\frac{R}{\tau}+\frac{1}{\tau}\gamma p(t) /M\,.
\end{equation}
Then non-Markovian equations \req{lange3} turn into the set of Markovian equations
\belar{lange4}
dq/dt&=&p(t)/M ,\qquad\qquad\qquad\qquad\qquad\qquad\nonumber\\
dp/dt&=&-[\partial E_{pot}/\partial q+R(t)] + \sqrt{ D}\zeta\,,\nonumber\\
dR/dt&=&- R / \tau +(\gamma p/M)/\tau\,,\\
d\zeta/dt&=&- \zeta / \tau +\xi/\tau\,.\nonumber
\end{eqnarray}
That is this set of equations that we will solve numerically below.

The integration of second Langevin equation \req{lange4} results in
\bel{pt6}
p(t+\Delta t)=p(t)-[\partial E_{pot}/\partial q+R(t)] \Delta t + \sqrt{D} \int_{t}^{t+\Delta t}\zeta(s) ds
\end{equation}

For the evaluation of the integral on the right we will
 write down the solution of  equation \req{lange4} for $\zeta(t)$   in two ways
\belar{pt17}
\zeta(t+\Delta t)&=\zeta(t)-\frac{1}{\tau}\int_t^{t+\Delta t} \zeta(s) ds+\frac{1}{\tau}\int_t^{t+\Delta t} \xi(s) ds\,,\nonumber\\
\zeta(t+\Delta t)&=\zeta(t)e^{-\Delta t/\tau}+\frac{1}{\tau}\int_t^{t+\Delta t}
e^{-\frac{t+\Delta t-s}{\tau}} \xi(s) ds\,.
\end{eqnarray}
By subtracting the second line from the first one will find
\belar{zint}
\int_t^{t+\Delta t} \zeta(s) ds=\tau[1-e^{-\Delta t/\tau}]\zeta(t) \,+\\
+\int_t^{t+\Delta t}[1-e^{-\frac{t+\Delta t-s}{\tau}}] \xi(s) ds\,.\nonumber
\end{eqnarray}
This relation is exact.
The integral in \req{zint} can be calculated in the same way as it was done above  for $\int_t^{t+\Delta t} \xi(s) ds$.
Then one will get
\belar{zinta}
\int_t^{t+\Delta t} \zeta(s) ds=\tau[1-e^{-\Delta t/\tau}]\zeta(t)\\
+\sqrt{\Delta t-2\tau(1-e^{-\Delta t/\tau})+\tau(1-e^{-2\Delta t/\tau})/2}\,\,\omega_1\,.\nonumber
\end{eqnarray}
The main order term of \req{zinta} in $\Delta t/\tau$ is:
\belar{limit}
\int_t^{t+\Delta t} \zeta(s) ds \approx \zeta(t)\Delta t\,, \qquad \text{for} \qquad\Delta t << \tau\,.
\end{eqnarray}
It looks like the colored noise can be integrated like an analytical function.

For the integration of \req{drdt} one can use the formal solution,
\bel{Rdt1}
R(t+\Delta t)=R(t)e^{-\frac{\Delta t}{\tau}}+\frac{1}{\tau}\int_t^{t+\Delta t}dt^{\pr}e^{-\frac{t+\Delta t-t^{\pr}}{\tau}}\gamma p(t^{\pr}) /M\,.
\end{equation}
The integral in \req{Rdt1} can be calculated assuming that $p(t^{\pr})$ does not change much on the time interval $\Delta t$. Then
\bel{Rdt2}
R(t+\Delta t)=R(t)e^{-\frac{\Delta t}{\tau}}+[\gamma p(t) /M](1-e^{-\frac{\Delta t}{\tau}})\,.
\end{equation}
For $\tau\to 0$ one easily gets the Markovian limit, $R(t)=\gamma p(t) /M$.

Finally, for the fourth equation of \req{lange4} one can use the second line of \req{pt17}.
The integral in \req{pt17}
\bel{pt17a}
y(t)=\frac{1}{\tau}\int_t^{t+\Delta t}e^{-{\frac{t+\Delta t-s}{\tau}}}\xi(s) ds
\end{equation}
is a sum of normally distributed random numbers,
and thereby is itself a normally distributed random number.
The random number is completely defined by its average number and the variance. From Eqs. (\ref{rand}, \ref{pt17a}) it is obvious that the average value is zero, $\langle y(t) \rangle=0$. The variance can be found by use of Eq. \req{rand},
\bel{varyt}
\langle y^2(t) \rangle=[1-e^{-2\Delta t/\tau}]/\tau\,.
\end{equation}
Hence, $y(t)$ can be represented as
\bel{pt17d}
y(t)=\sqrt{\frac{1}{2\tau}}\sqrt{1-e^{-2\Delta t/\tau}}\,\omega_1\,,
\end{equation}
and Eq. \req{pt17} turns into
\bel{pt18}
\zeta(t+\Delta t) = \zeta(t)e^{-\Delta t/\tau}+\frac{1}{\sqrt{2\tau}}\sqrt{1-e^{-2\Delta t/\tau}}\,\omega_1\,.
\end{equation}

Below we investigate separately the influence of the memory effects in the friction and random force on the fission width.

\begin{figure}[ht]
\centering
\includegraphics[width=0.35\textwidth]{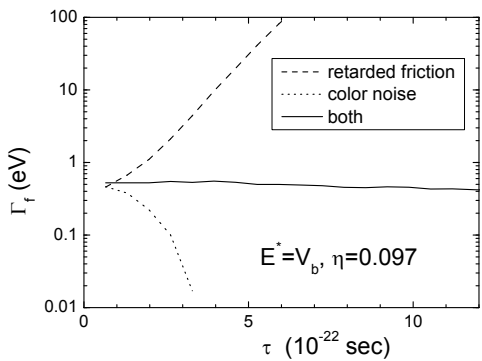}
\caption{The variation of the fission width at $E=V_b$ due to the memory effects in the friction force alone (dash), in the random force alone (dot), and both in friction and random forces (solid).
Parameters of calculations: $E^*=V_b, \Delta t=5\cdot 10^{-24} sec, t_{max}=6\cdot  10^{-20} sec, N_{tr}=1 000 000$.}
\label{retarded}
\end{figure}
The fission width calculated with Eq.\req{Rdt2} and
\bel{pt66}
p(t+\Delta t)=p(t)-\bigg[\frac{\partial E_{pot}}{\partial q}+R(t)\bigg] \Delta t + \sqrt{D} \int_{t}^{t+\Delta t}\xi(s) ds\,,
\end{equation}
for $E^*=V_b$ is shown by red line in Fig.~\ref{retarded}.
As one can see, the account of memory effects in the friction force makes $\Gamma_f$ much larger.

The blue line in Fig.~\ref{retarded} is the calculation with memory effects in the random force \req{zinta} and Markovian friction coefficient. As one can see, the memory effects in the random force make fission width much smaller.
The fission width calculated with (\ref{pt6}, \ref{zinta}, \ref{Rdt2}) (memory effects both in friction and random forces) is shown by black line in Fig.~\ref{retarded}. The memory effects in the friction and random forces almost cancel each other, and the variation of the black curve with $\tau$ is very small.
\begin{figure*}[!htb]
\centering
\includegraphics[width=0.85\textwidth]{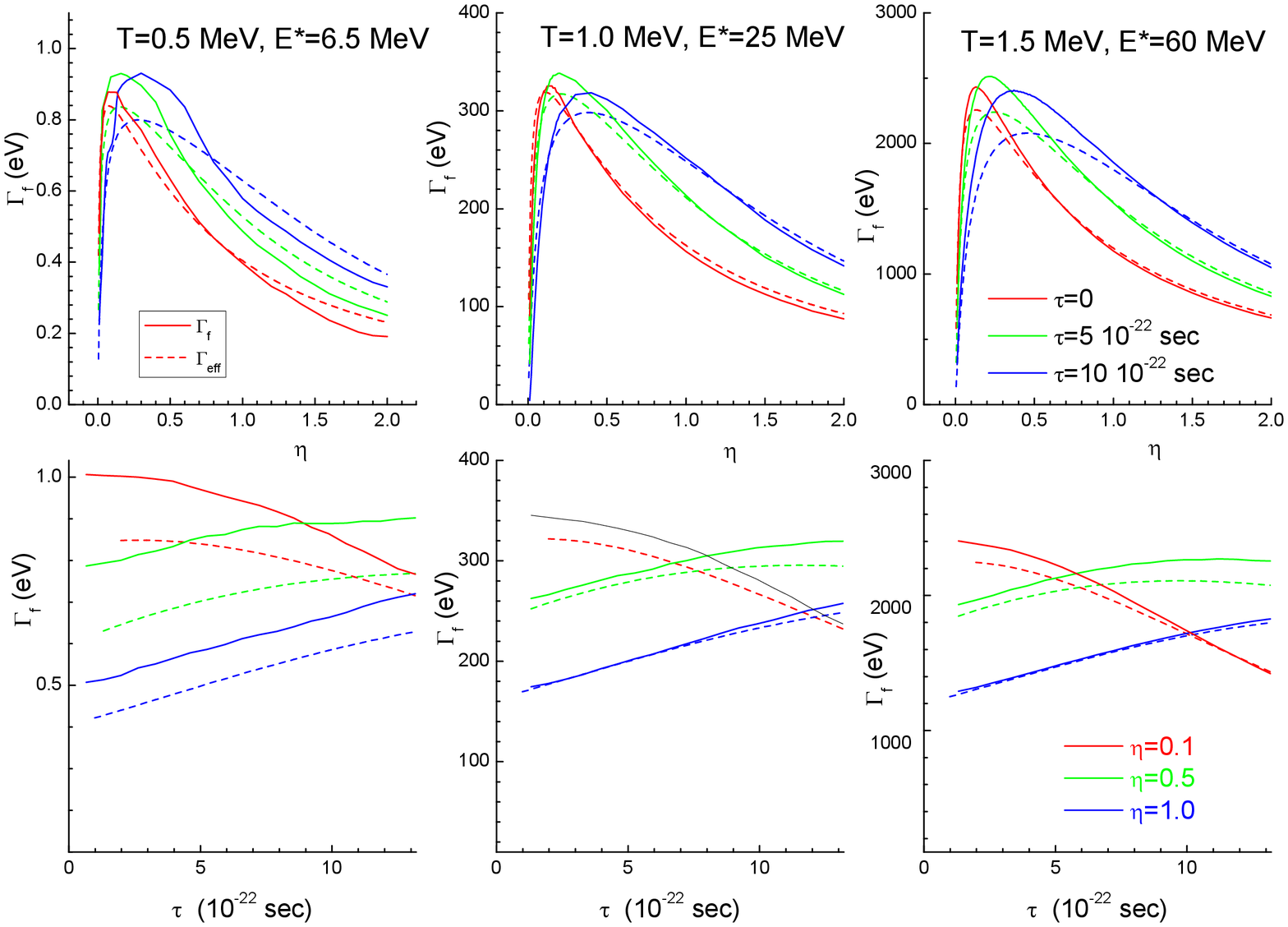}
\caption{(Color online) Top: The dependence of the fission width $\Gamma_f$ on the damping parameter $\eta$ for few values of the relaxation $\tau$=0, $\tau=5\cdot 10^{-22}$ sec,  $\tau=10^{-21}$ sec and temperatures, T=0.5, 1.0 and 1.5 MeV. Parameters of the calculations,
$\Delta t=5\cdot 10^{-24} sec, t_{max}=6\cdot  10^{-20} sec, N_{tr}=1 000 000$.
Bottom: The dependence of the fission width $\Gamma_f$ on the relaxation time $\tau$  for few values of damping parameter $\eta$, $\eta$=0.1, 0.5 and 1.0 .}
\label{graph1}
\end{figure*}

The results of the detailed investigations of memory effects on the fission width are shown below in Fig.~\ref{graph1}. In the top part of Fig.~\ref{graph1} the fission width $\Gamma_f$ is shown as a function of the damping parameter $\eta$ for few values of the relaxation time both for small and large  excitation energies (temperatures), $T$=0.5, 1.0 and 1.5 MeV.
Since at present the theoretical estimates for $\tau$ vary in very board region, besides $\tau=0$, we  choose in calculations below the two values of $\tau$ close to those used in \cite{adeev2008}, namely $\tau=5\cdot 10^{-22}$ sec and $\tau=10^{-21}$ sec. The damping parameter $\eta$ here is not related to the friction force by \req{GammaK}, but is considered as a free parameter. Consequently, in Langevin calculations the friction parameter $\gamma$ was related to $\eta$ by $\gamma=2 M \omega_0 \eta$.

The results of Langevin calculations are shown in Fig.~\ref{graph1} by solid lines. As one can see, the dependence of $\Gamma_f$ on $\eta$ and $\tau$ is rather complicated. In low viscosity region the fission width $\Gamma_f$ {\it grows} as function of $\eta$ and {\it decreases} as function of $\tau$. In high viscosity region the tendency is opposite,  the fission width $\Gamma_f$ {\it falls down} as function of $\eta$ and {\it increases} as function of $\tau$. Such dependence is common both for small and large excitation energies. In the intermediate region $\Gamma_f$ does not depend much both on $\eta$ and $\tau$. The turnover point in $\eta$ depends on the value of $\tau$. It varies from $\eta\approx 0.1$ for $\tau=0$ to $\eta\approx 0.5$ for $\tau=10^{-21}$ sec.

In the bottom part of Fig.~\ref{graph1} the fission width $\Gamma_f$ is shown as function of the relaxation time $\tau$ for few fixed values of the damping parameter $\eta$, namely for $\eta$= 0.1, 0.5 and 1.0. The bottom part of Fig.~\ref{graph1} confirms the above conclusion: In low viscosity region the fission width $\Gamma_f$ {\it grows} as function of $\eta$ and {\it decreases} as function of $\tau$. In high viscosity region the tendency is opposite,  the fission width $\Gamma_f$ {\it falls down} as function of $\eta$ and {\it increases} as function of $\tau$.

For the comparison, in Fig.~\ref{graph1} we show by dash lines the available analytical approximation for $\Gamma_f$. By now there exist the generalization of  Kramers low and high viscosity limits for the case of finite relaxation time $\tau$.

For large damping $\eta\gg 1$, the fission width is well described by the approximation
\cite{grote,abe-san,lallouet}
\bel{gammahvt}
\Gamma_{HV}(\tau)=\frac{\hbar\lambda}{2\pi}e^{-V_b/T}\,.
\end{equation}
where $\lambda$ is the largest positive solution of the secular equation
\bel{laluet}
\lambda^3+\frac{\lambda^2}{\tau}+\biggl(\frac{\beta}{\tau}-\omega^2\biggr)\lambda-\frac{\omega^2}{\tau}=0
\end{equation}
For small friction there exists a modification of Kramers low-viscosity limit \cite{fonseca}

\bel{gammalvt}
\Gamma_{LV}(\tau)=\Gamma_{LV}/(1+\omega_0^2\tau^2)\,.
\end{equation}
For the interpolation between high viscosity and low viscosity limits one often uses the popular expression \cite{hanggi}
\bel{gammaeff}
\frac{1}{\Gamma_{eff}}=\frac{1}{\Gamma_{LV}}+\frac{1}{\Gamma_{HV}}\,.
\end{equation}
In limits $\eta<< 1$ or $\eta>> 1$ $\Gamma_{eff}$ approaches $\Gamma_{LV}$ \req{gammalvt}, or $\Gamma_{HV}$  \req{gammahvt}, correspondingly.

\begin{figure}[htp]
\includegraphics[width=0.35\textwidth]{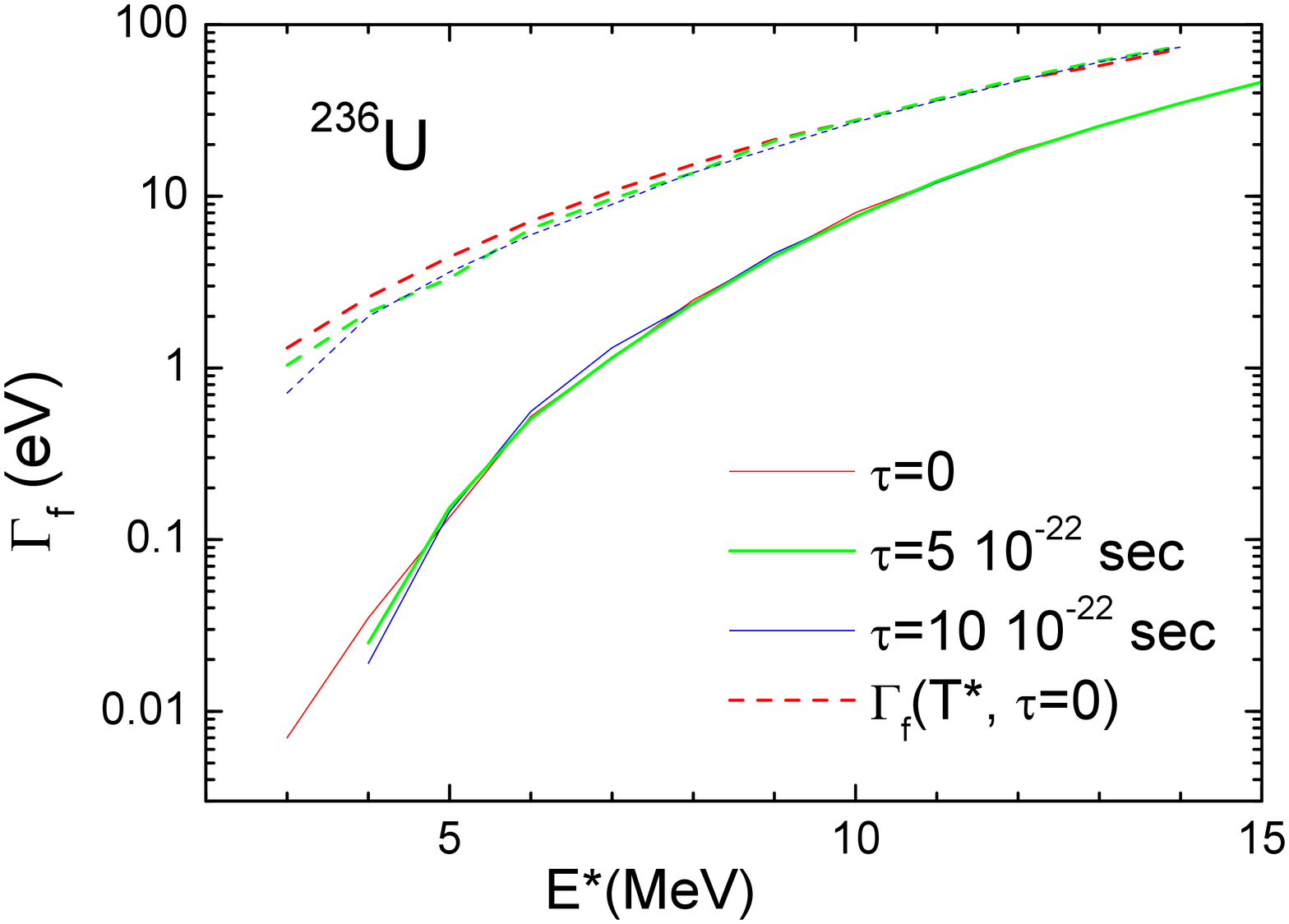}
\centering
\caption{(Color online) The dependence of the fission width $\Gamma_{f}$ \req{Gf} on the excitation energy for the transport coefficients (\ref{epot}, \ref{massa}, \ref{fric}). Solid: red line - fission width $\Gamma_{f}$ \req{Gf} for $\tau=0$, green - for $\tau=5\cdot 10^{-22}$ sec, blue - for $\tau=10^{-21}$ sec.
Dash lines - the fission width $\Gamma_{f}$ caculated with the effective temperature T* \protect\req{teff1} for the same values of $\tau$.
}
\label{gamma_ex}
\end{figure}

The results of Langevin calculations look very similar to the analytical estimate of fission width \req{gammaeff}. As one can see, the results of Langevin calculations approach analytical estimate \req{gammaeff} both in low and high viscosity limits. For the intermediate values of $\eta$ the Langevin results are somewhat larger than that given by the interpolation formula \req{gammaeff}.

Since the main interest in our work are the memory effects at small excitation energies, we show in Fig.~\ref{gamma_ex} the  dependence of the fission width $\Gamma_{\tau}$ calculated with (\ref{pt6}, \ref{zinta}, \ref{Rdt2}) on the excitation energy $E^*$, $4 \mev \leq E^*\leq 12 \mev$ for few values of the memory time,  $\tau=0, \tau=0.5\cdot 10^{-22}$ sec and $10^{-21}$ sec. 
In these calculations the mass and friction parameters were chosen according to Eqs. \req{massa}, \req{fric}, i.e. were fitted to the results obtained within the linear response theory \cite{hiry}. For these values of mass and friction the damping parameter vary within the limits $0.075\leq \eta \leq  0.3$, i.e. belong mainly to the intermediate damping region.
As it follows from above, in the intermediate damping region the dependence of $\Gamma_f$ on $\tau$ is very week, which is confirmed by the numerical results in Fig.~\ref{gamma_ex}. Only at $E^*$= 4 MeV ($\eta$=0.075) one can see some dependence of $\Gamma_f$ on $\tau$. The fission width $\Gamma_f$ here is getting smaller with growing $\tau$.
\section{Effective temperature}
The only quantity in the Langevin approach \req{lange4} that was not modified so far is the diffusion coefficient $D$. In principle, the diffusion coefficient also can contain the quantum effects. As it was shown by H.Hofmann and D.~Kiderlen \cite{hofkid} in the quantum regime the classical Einstein relation for the diffusion coefficient $D=\gamma T$ should be modified to
\bel{efftemp}
D=\gamma T \to D^*=\gamma T^*\,
\end{equation}
with
\bel{teff1}
T^\ast(\omega_0) =(\hbar \omega_0 /2)\,\,\coth \,\,(\hbar \omega_0 /2T)\,\,.
\end{equation}
for the positive stiffness. The parameter $\omega_0$ is the local frequency of collective motion \cite{hofkid}. The minimal value of the effective temperature T$^{\ast}$ is given by $\hbar \omega_0/2$. For the negative stiffness Eq. \req{teff1} takes the form \cite{hofbook}
\bel{teff2}
T^\ast(i\omega_b) =(\hbar \omega_b /2)\,\,\cot \,\,(\hbar \omega_b /2T)\,\,.
\end{equation}
Here $\omega_b$ is the frequency of collective motion around the fission barrier, $\omega_b^2=|C|/M$.
Since $T^*$ should be positive, the application of \req{teff2} will break down at a critical temperature $T_c$,
\bel{Tcrit}
T_c =\hbar \omega_b /\pi.
\end{equation}
Now, we simply replace the diffusion coefficient $D$ in \req{lange4} by $D^\ast$. The fission width calculated with the effective temperature (\ref{efftemp}-\ref{teff2}) for $\tau=0$ is shown in Fig.~\ref{gamma_ex} by the dashed lines.  As one can see from Fig.~\ref{gamma_ex}, the effect of effective temperature \req{efftemp} is huge. Depending on the excitation energy, $\Gamma_f$ is getting larger by up to two orders of magnitude, compared with the value calculated with $D=\gamma~T$.

Fig.~\ref{gamma_ex} looks very similar to the Fig.~2 of \cite{washiyama} for the probability of compound nucleus formation. It was stressed in \cite{washiyama} that the
quantum effects increase the compound nucleus formation probability at low excitation
energies. At $T=0.5 \mev$ the quantum enhancement of probability of compound nucleus formation in \cite{washiyama} and the enhancement of fission width in present work ($T=0.5 \mev$ corresponds to $E^*= 6 \mev $) is approximately the same - one order of magnitude. But the effect of $T^*$ on the fission width increases rapidly at a lower temperature (excitation energies).
At $E^*=V_b/2 ~(T=0.335 \mev)$ the fission width calculated with $T$ or $T^*$ differ by two orders of magnitude. It is difficult to believe that such a huge effect makes sense.

The reason for such an effect may be related to some inconsistency of the present approach. We use the very simple approximation for friction and inertia - constant deformation independent values - and at the same time, take into account the quantum effects in the diffusion coefficient. It would be more consistent to consider all transport coefficients in the same approach, say within the linear response theory.

The excitation energies shown in Fig.~\ref{gamma_ex} are restricted by the condition \req{Tcrit}. To go to smaller $E^*$, one may try to avoid using the effective temperature. For this let us note that the diffusion coefficient \req{efftemp} can be identically written as
\begin{equation}
D^*(\omega)=\gamma(\omega) T^*(\omega)= \frac{\chi^{\prime\prime}(\omega)}{\omega}\frac{\hbar\omega}{2}\coth{\frac{\hbar\omega}{2T}}=\frac{1}{2}\Psi^{\prime\prime}(\omega)
\label{DPsi}.
\end{equation}
Here for the friction coefficient we used the common relation from the linear response theory, $\gamma(\omega)=\chi^{\dpr}(\omega)/\omega$ and the relation between the imaginary parts of the response $\chi^{\prime\prime}(\omega)$ and the correlation $\Psi^{\prime\prime}(\omega)$ functions
\begin{equation}
\Psi^{\prime\prime}(\omega)=\hbar \chi^{\prime\prime}(\omega) \coth(\hbar\omega/2T)\,.
\label{FDT}
\end{equation}
The correlation function $\Psi^{\prime\prime}(\omega)$ can be calculated within the linear response theory for any shape, for any temperature, like it was done in \cite{ivahof}, even with pairing effects taken into account. In this way, one could avoid the use of effective temperature and the restriction \req{Tcrit}.

The calculations within the Langevin approach with the microscopic transport coefficients will be the subject of the next studies.
\bigskip
\section{Summary}

$\bullet$ We have investigated the role of memory effects on the fission width within the Langevin approach with a schematic one-dimensional model with the potential energy given by the two-parabolic potential. The deformation-independent mass and friction parameters were fitted to the results, obtained within the microscopic linear response theory.

$\bullet$The use of a simple model gives a chance to examine
the dependence of fission width $\Gamma_f$ on the memory time and the excitation energy at low temperatures where the quantum fluctuations play an important role.

$\bullet$It turns out that the dependence of $\Gamma_f$ on the relaxation time $\tau$ is very sensitive to the damping parameter $\eta$. In the low viscosity region the fission width $\Gamma_f$ {\it grows} as function of $\eta$ and {\it decreases} as function of $\tau$. In high viscosity region the tendency is opposite.  The fission width $\Gamma_f$ {\it decreases} as function of $\eta$ and {\it increases} as function of $\tau$. Such dependence is common both for small and large excitation energies. The turnover point in $\eta$ depends on the value of $\tau$. It varies from $\eta\approx 0.1$ for $\tau=0$ till $\eta\approx 0.5$ for $\tau=10^{-21}$ sec.

$\bullet$ For the excitation energies around the fission barrier the dependence of $\Gamma_f$ on the relaxation time is negligibly small. This conclusion may depend on the model for friction force, used in the calculations.

$\bullet$ The replacement of the temperature by the effective temperature in the diffusion coefficient increases the fission width at low excitation energies up to two orders of magnitude. This effect seems unreasonably too big.

$\bullet$ For the further investigation of the role of effective temperature it would be worth carrying out the Langevin calculations with all transport coefficients defined within the microscopic approach, say, within the linear response theory.

\bigskip

\begin{acknowledgments}
 This study was supported in part by the program "Support for the development of priority areas of
scientific research" of the National Academy of Sciences of Ukraine, Grant No. 0120U100434. The authors appreciate very much the fruitful discussions with Prof. Y. Abe and Prof. K. Hagino.

\end{acknowledgments}

\end{document}